\documentclass[pra,aps,showpacs,twocolumn]{revtex4}
\usepackage{graphicx}
\usepackage{latexsym}
\usepackage{amsmath}
\usepackage[dvipdfm]{hyperref}

\input{tcilatex}
\begin{document}

\title{Quantifying coherence of Gaussian states}
\author{Jianwei Xu}
\email{xxujianwei@nwafu.edu.cn}
\affiliation{College of Science, Northwest A\&F University, Yangling, Shaanxi 712100,
China}
\date{\today}

\begin{abstract}
Coherence arises from the superposition principle and plays a key
role in quantum mechanics. Recently, Baumgratz et al. [T. Baumgratz, M.
Cramer, and M. B. Plenio, Phys. Rev. Lett. 113, 140401 (2014)] established a
rigorous framework for quantifying the coherence of finite dimensional
quantum states. In this work we provide a framework for quantifying the
coherence of Gaussian states and explicitly give a coherence measure based
on the relative entropy.
\end{abstract}

\pacs{03.65.Ud, 03.67.Mn, 03.65.Aa}
\maketitle

\bigskip

\section{Introduction}
Coherence is a basic feature in quantum mechanics, it is a common necessary
condition for both entanglement and other types of quantum correlations.
Many works have been undertaken to theoretically formulate quantum coherence
\cite{Glauber1963,Sudarshan1963,Luo2005,Aberg2006,Monras2014,Aberg2014,Girolami2014}, but up to now there has
been no well-accepted efficient method for quantifying coherence.
Recently, Baumgratz et al. established a comprehensive framework of
coherence quantification \cite{Baumgratz2014}, by
which coherence is considered to be a resource that can be characterized,
quantified, and manipulated in a manner similar to quantum entanglement
\cite{Bennett1996,Vedral1998,Plenio2007,Horodecki2009}. This seminal work has triggered the community's interest of other proper measures for coherence \cite{Streltsov2015,Shao2015}, the freezing phenomenon \cite{Bromley2015}, and some further developments \cite{Xi2015,Pires2015,Bera2015,Singh2015,Yadin2015,Yuan2015,Yao2015}. Ref.\cite{Du2015} revealed the condition of coherence
transformations for pure states, and then built a universal method for
quantifying the coherence of mixed states via the convex roof scheme.

All above results for quantifying quantum coherence are implicitly
assumed the finite dimensional setting, which is neither necessary nor
desirable. In relevant physical situations such as quantum optics, it must
require the quantum states in infinite dimensional systems, especially the
Gaussian states \cite{Braunstein2005,Wang2007,Weedbrook2012}. In this work we investigate the
quantification of coherence for Gaussian states.

This work is organized as follows. In section II, we discuss the necessary conditions any measure of coherence for Gaussian
states should satisfy.  In section III, we prove that the incoherent
one-mode Gaussian states are just thermal states. In section IV, we
determine the structure of one-mode incoherent Gaussian operations. In
section V, we explicitly provide a coherence measure for one-mode Gaussian states
based on the relative entropy. In section VI we consider the multi-mode
case. Section VII is a brief summary.

\section{How to quantify the coherence of Gaussian states}
A state $\rho $ (finite or infinite dimensional) is said to be incoherent if
it is diagonal when expressed it in a fixed orthonormal basis. We denote the
set of all incoherent states by $\mathcal{I}$. A quantum map is called
incoherent operation (ICPTP) if it is completely positive, trace-preserving,
and maps any incoherent states into incoherent states. Ref.\cite{Baumgratz2014} presented the postulates that any proper
measure of the coherence $C(\rho )$ for finite-dimensional state $\rho $
must satisfy as follows.

$(C1)$ $C(\rho )\geq 0$ and $C(\rho )=0$ iff $\rho \in \mathcal{I}$.

$(C2a)$ Monotonicity under all incoherent completely positive and
trace-preserving (ICPTP) maps: $C(\rho )\geq C(ICPTP(\rho )).$

$(C2b)$ Monotonicity for average coherence under subselection based on
measurement outcomes: $C(\rho )\geq \sum_{n}p_{n}C(\rho _{n})$,
where $\rho _{n}=K_{n}\rho K_{n}^{+}/p_{n}$ and $p_{n}=tr(K_{n}\rho
K_{n}^{+})$ for all $n$, $\sum_{n}K_{n}^{+}K_{n}=I$, $K_{n}\mathcal{I}%
K_{n}^{+}\subset \mathcal{I}$, with $+$ the adjoint and $I$ identity
operator.

$(C3)$ Nonincreasing under the mixing of quantum states:
$\sum_{n}p_{n}C(\rho _{n})\geq C(\sum_{n}p_{n}\rho _{n}).$

Note that $(C2b)$ and $(C3)$ together imply $C(2a)$.

For the case of Gaussian states, we adopt $(C1)$ and $C(2a)$ as necessary
conditions that any coherence measure should satisfy, while give up $C(2b)$
and $C(3)$. Gaussian states do not form a convex set, then it seems hard to
establish the counterparts of $C(2b)$ and $C(3)$.

\section{Incoherent states of one-mode Gaussian states}
In this section, we find out the incoherent states of one-mode Gaussian
states. We first note that, coherence is basis dependent, so wherever we
talk about coherence we must be clear which basis is presupposed.

\textbf{\emph{Theorem 1.}} For fixed orthonormal basis $\{|n\rangle \}_{n=0}^{\infty }$, a
one-mode Gaussian state is diagonal iff it is a thermal state.

\textbf{Proof.} A state $\rho $ is called Gaussian if its characteristic function $%
\chi (\rho ,\lambda )=tr[\rho D(\lambda )]$ is of the form
\begin{eqnarray}
\chi (\rho ,\lambda )=\exp \{-\frac{1}{2}(x_{\lambda },y_{\lambda })\Omega
V\Omega ^{t}\binom{x_{\lambda }}{y_{\lambda }} \nonumber  \\
-i[\Omega \binom{d_{1}}{d_{2}}%
]^{t}\binom{x_{\lambda }}{y_{\lambda }}\},
\end{eqnarray}
where $D(\lambda )=e^{\lambda a^{+}-\lambda ^{\ast }a}$ is the displacement
operator, $a$, $a^{+}$ are creator operator and annihilation operator, $%
x_{\lambda }$ and $y_{\lambda }$ are the real and imaginary parts of $%
\lambda $, $\Omega =\left(%
\begin{array}{cc}
0 & 1 \\
-1 & 0%
\end{array}\right)
,d=(d_{1},d_{2})^{t}$ with $d_{1}$ and $d_{2}$ real numbers, $t$ denotes
transpose, $V=\left(
\begin{array}{cc}
V_{11} & V_{12} \\
V_{21} & V_{22}%
\end{array}\right)
$ is real symmetric positive and satisfies $V+i\Omega \geq 0$. A Gaussian
state $\rho $ is fully described by the covariance matrix $V$ and the
displacement vector $d$. $detV\geq 1$ and $detV=1$ iff $\rho $ is pure. We
write a Gaussian state as $\rho (V,d).$

A state is called thermal if it has the form
\begin{eqnarray}
\rho _{th}(\overline{n})=\sum_{n=0}^{\infty }\frac{\overline{n}^{n}}{(%
\overline{n}+1)^{n+1}}|n\rangle \langle n|,
\end{eqnarray}
where $\overline{n}=tr[a^{+}a\rho _{th}(\overline{n})]\geq 0$ is the mean
number.

It is easy to check that the characteristic function of the thermal state $%
\rho _{th}(\overline{n})$ is Gaussian with covariance matrix $(2\overline{n}%
+1)I$ and zero displacement vector. Hence we only need to prove that the
diagonal Gaussian states must be thermal states. To this aim, we calculate
the elements $\rho _{mn}=$ $\langle m|\rho |n\rangle $ from Eq.(1) and its
inverse relation
\begin{eqnarray}
\rho =\int \frac{d^{2}\lambda }{\pi }\chi (\rho ,\lambda )D(-\lambda ),
\end{eqnarray}
where $d^{2}\lambda =dx_{\lambda }dy_{\lambda }$ and $\int =\int_{-\infty
}^{\infty }$. We thus have
\begin{eqnarray}
\langle m|\rho |n\rangle =\int \frac{d^{2}\lambda }{\pi }\chi (\rho
,\lambda )\langle m|D(-\lambda )|n\rangle ,
 \\
\langle m|D(-\lambda )|n\rangle \ \ \ \ \ \ \ \ \ \ \ \ \ \ \ \ \ \ \ \ \ \ \ \ \ \ \ \ \ \ \ \ \ \      \nonumber \\
=\int \int \frac{d^{2}\alpha }{\pi }\frac{%
d^{2}\beta }{\pi }\langle m|\alpha \rangle \langle \alpha |D(-\lambda
)|\beta \rangle \langle \beta |n\rangle ,
 \\
\langle \alpha |D(-\lambda )|\beta \rangle =\langle 0|D(-\alpha )D(-\lambda
)D(\beta )|0\rangle ,
\end{eqnarray}
where $|\alpha \rangle =e^{-\frac{|\alpha |^{2}}{2}}\sum_{n=0}^{\infty }%
\frac{\alpha ^{n}}{\sqrt{n!}}|n\rangle $, $|\beta \rangle =e^{-\frac{|\beta
|^{2}}{2}}\sum_{n=0}^{\infty }\frac{\beta ^{n}}{\sqrt{n!}}|n\rangle $ are
coherent states. Using the formula
\begin{eqnarray}
D(\alpha )D(\beta )=e^{\frac{\alpha \beta ^{\ast }-\alpha ^{\ast }\beta }{2}%
}D(\alpha +\beta ),
\end{eqnarray}
and after direct algebras, we get
\begin{eqnarray}
\langle m|\rho |n\rangle =\int \int \int \frac{d^{2}\lambda }{\pi }\frac{%
d^{2}\alpha }{\pi }\frac{d^{2}\beta }{\pi }\frac{\alpha ^{m}\beta ^{\ast n}}{%
\sqrt{m!n!}}\exp b,  \ \ \ \
\\
b=-\alpha ^{\ast }\lambda +\alpha ^{\ast }\beta +\lambda ^{\ast }\beta
-|\alpha |^{2}-|\beta |^{2}-\frac{1}{2}|\lambda |^{2}-\chi (\rho ,\lambda ).
\end{eqnarray}
Eq.(8) is somewhat similar to the results of Refs.\cite{Isserlis1918,Wick1950,Withers1985}, but in fact not the same thing.
To calculate Eq.(8), we introduce the integration
\begin{eqnarray}
J=\int \int \int d^{2}\lambda d^{2}\alpha d^{2}\beta \exp \{b+u\alpha
+v\beta ^{\ast }\},
\end{eqnarray}
where $u$, $v$ are real numbers. As a result,
\begin{eqnarray}
\rho _{mn}=\frac{1}{\pi ^{3}}(\frac{\partial ^{m}}{\partial u^{m}}\frac{%
\partial ^{n}}{\partial v^{n}}J)_{u=v=0}.
\end{eqnarray}
$J$ is a Gaussian integration we can calculate as follows.

Write $b+u\alpha +v\beta ^{\ast }$ as
\begin{eqnarray}
b+u\alpha +v\beta ^{\ast }=-\frac{1}{2}(x_{\alpha },y_{\alpha },x_{\beta
},y_{\beta },x_{\lambda },y_{\lambda })A  \ \ \ \ \ \ \ \      \nonumber \\
\cdot(x_{\alpha },y_{\alpha },x_{\beta
},y_{\beta },x_{\lambda },y_{\lambda })^{t}+B(x_{\alpha },y_{\alpha
},x_{\beta },y_{\beta },x_{\lambda },y_{\lambda })^{t},
\end{eqnarray}
where $A$ is a $6\times 6$ symmetric complex matrix, $B$ is a $6\times 1$
row complex vector,
\begin{eqnarray}
A=\left(
\begin{array}{cccccc}
2 & 0 & -1 & -i & 1 & i \\
0 & 2 & i & -1 & -i & 1 \\
-1 & i & 2 & 0 & -1 & i \\
-i & -1 & 0 & 2 & -i & -1 \\
1 & -i & -1 & -i & 1+V_{22} & -V_{12} \\
i & 1 & i & -1 & -V_{12} & 1+V_{11}%
\end{array} \right),
\\
detA=16(\det V+V_{11}+V_{22}+1)>0, \ \ \ \ \ \ \ \
\\
B=(u,iu,v,-iv,-id_{2},id_{1}).   \ \ \ \ \ \ \ \ \ \ \ \ \ \ \ \ \ \ \ \ \
\end{eqnarray}
Apply the Gaussian integration formula we get
\begin{eqnarray}
J=\frac{(2\pi )^{3}}{\sqrt{\det A}}\exp [\frac{1}{2}BA^{-1}B^{t}].
\end{eqnarray}
Let
\begin{eqnarray}
\xi =\frac{1}{2}BA^{-1}B^{t}=\frac{1}{2}%
[(u,v)B_{2}(u,v)^{t}+B_{1}(u,v)^{t}+B_{0}], \ \
\end{eqnarray}
where
\begin{eqnarray}
B_{2}=\left(
\begin{array}{cc}
\frac{V_{11}-V_{22}+2iV_{12}}{1+V_{11}+V_{22}+V_{11}V_{22}-V_{12}^{2}} &
\frac{V_{11}V_{22}-V_{12}^{2}-1}{1+V_{11}+V_{22}+V_{11}V_{22}-V_{12}^{2}} \\
\frac{V_{11}V_{22}-V_{12}^{2}-1}{1+V_{11}+V_{22}+V_{11}V_{22}-V_{12}^{2}} &
\frac{V_{11}-V_{22}-2iV_{12}}{1+V_{11}+V_{22}+V_{11}V_{22}-V_{12}^{2}}%
\end{array}\right), \nonumber
\\
B_{1}=2(\frac{(1-iV_{12}+V_{22})d_{1}+i(1+V_{11}+iV_{12})d_{2}}{%
1+V_{11}+V_{22}+V_{11}V_{22}-V_{12}^{2}}, \ \ \ \ \ \ \ \ \   \nonumber  \\
\frac{%
(1+iV_{12}+V_{22})d_{1}-i(1+V_{11}-iV_{12})d_{2}}{%
1+V_{11}+V_{22}+V_{11}V_{22}-V_{12}^{2}})  \ \ \ \ \ \ \ \ \  \nonumber
\\
B_{0}=-\frac{(1+V_{22})d_{1}^{2}-2V_{12}d_{1}d_{2}+(1+V_{11})d_{2}^{2}}{%
1+V_{11}+V_{22}+V_{11}V_{22}-V_{12}^{2}}. \ \ \ \ \ \ \ \ \ \ \ \ \       \nonumber
\end{eqnarray}
Introduce the symbols
\begin{eqnarray}
J_{k_{1}k_{2}...k_{m}}=\frac{\partial ^{m}J}{\partial k_{1}\partial
k_{2}...\partial k_{m}},
\\
\xi _{k_{1}k_{2}...k_{m}}=\frac{\partial ^{m}\xi }{\partial k_{1}\partial
k_{2}...\partial k_{m}},
\end{eqnarray}
where $k_{1},k_{2},...k_{m}\in \{u,v\}$, $m\in \{1,2,3,...\},$and let $J(0)$%
, $J_{k_{1}k_{2}...k_{m}}(0)$, $\xi (0)$, $\xi _{k_{1}k_{2}...k_{m}}(0)$
represent the corresponding values when $u=v=0$. Direct calculations show that
\begin{eqnarray}
\xi _{k_{1}}(0)=\frac{1}{2}(B_{1})_{k_{1}},  \ \ \ \ \ \ \ \ \ \ \ \ \ \ \ \ \ \ \ \ \ \ \ \ \ \ \ \ \ \ \ \ \ \
\\
\xi _{k_{1}k_{2}}(0)=(B_{2})_{k_{1}k_{2}},  \ \ \ \ \ \ \ \ \ \ \ \ \ \ \ \ \ \ \ \ \ \ \ \ \ \ \ \ \ \ \ \ \ \
\\
\xi _{k_{1}k_{2}...k_{m}}(0)=0 \text{\ when \ } m\geq 3, \ \ \ \ \ \ \ \ \ \ \ \ \ \ \ \ \ \ \ \ \ \ \ \ \
\\
J_{k_{1}}(0)=J(0)\xi _{k_{1}}(0), \ \ \ \ \ \ \ \ \ \ \ \ \ \ \ \ \ \ \ \ \ \ \ \ \ \ \ \ \ \ \
\\
J_{k_{1}k_{2}}(0)=J(0)[\xi _{k_{1}k_{2}}(0)+\xi _{k_{1}}(0)\xi
_{k_{2}}(0)],  \ \ \ \ \ \ \ \
\\
J_{k_{1}k_{2}k_{3}}(0)=J(0)[\xi _{k_{1}}(0)\xi _{k_{2}k_{3}}(0)+\xi
_{k_{2}}(0)\xi _{k_{1}k_{3}}(0)   \nonumber  \\
  +\xi _{k_{3}}(0)\xi _{k_{1}k_{2}}(0)+\xi
_{k_{1}}(0)\xi _{k_{2}}(0)\xi _{k_{3}}(0)],...,
\\
J_{k_{1}k_{2}...k_{m}}(0)=J(0)\sum_{\sigma
(k_{1}k_{2}...k_{m})}\sum_{r=0}^{[\frac{m}{2}]}\xi _{k_{1}k_{2}}(0)\xi
_{k_{3}k_{4}}(0)   \nonumber \\
 ...\xi _{k_{2r-1}k_{2r}}(0)\xi _{k_{2r+1}}(0)...\xi
_{k_{m}}(0),
\end{eqnarray}
where $[\frac{m}{2}]=\frac{m}{2}$ when $m$ is even and $[\frac{m}{2}]=\frac{%
m-1}{2}$ when $m$ is odd, $\sigma (k_{1}k_{2}...k_{m})$ is any permutation
of $k_{1}k_{2}...k_{m}$, $\sum_{\sigma (k_{1}k_{2}...k_{m})}$ sums all
permutations of $k_{1}k_{2}...k_{m}$.

From Eqs.(11,26) we can calculate any $\rho _{mn}$ in principle.

If the Gaussian state $\rho $ is diagonal thus $\rho _{01}=\rho _{02}=0,$
\begin{eqnarray}
\rho _{01}=\frac{2^{3}}{\sqrt{\det A}}J(0)\frac{1}{2}(B_{1})_{2}=0%
\Rightarrow (B_{1})_{2}=0.
\\
\rho _{02}=\frac{2^{3}}{\sqrt{\det A}}J(0)[(B_{2})_{22}+\frac{1}{2}%
(B_{1})_{2}\frac{1}{2}(B_{1})_{2}]=0 \nonumber \\
\Rightarrow (B_{2})_{22}=0.
\end{eqnarray}
Similarly $\rho _{10}=\rho _{20}=0$ yield
\begin{eqnarray}
(B_{1})_{1}=0,  \\
(B_{2})_{11}=0.
\end{eqnarray}
Taking Eqs.(27-30) into $B_{2},B_{1},$ we get
\begin{eqnarray}
V_{11}-V_{22}=V_{12}=d_{1}=d_{2}=0,
\end{eqnarray}
hence $\rho $ is a thermal state. We then complete this proof.

\section{Incoherent operations of one-mode Gaussian states}
A one-mode Gaussian operation is described by $(T,N,\overline{d})$, it
performs on the Gaussian state $\rho (V,d)$ and get the Gaussian state with
the covariance matrix and displacement vector as \cite{Holevo2001}
\begin{eqnarray}
d\rightarrow Td+\overline{d},V\rightarrow TVT^{t}+N,
\end{eqnarray}
where $N$, $T$ are real matrices satisfying $N=N^{t}\geq 0,\det N\geq (\det
T-1)^{2}$.

We give the definition of incoherent operation for one-mode Gaussian states.
A one-mode Gaussian operation is called incoherent if it maps any incoherent
Gaussian state into incoherent Gaussian states. We denote the set of all
incoherent operations by $\mathcal{I}$. We now study the structure of
one-mode Gaussian incoherent operation.

Suppose one-mode Gaussian operation $(T,N,\overline{d})$ is incoherent, for
any incoherent state $\rho (rI,0)$ with $r\geq 1,$ after the action of $(T,N,%
\overline{d})$, $\rho (rI,0)$ becomes the incoherent state $\rho (sI,0)$
with $s\geq 1.$ From Eq.(32), we have
\begin{eqnarray}
\overline{d}=0,  \ \ \ \ \ \ \ \ \ \ \ \ \ \ \ \ \ \ \ \ \ \ \ \ \ \ \ \ \ \ \  \ \ \ \ \ \ \ \ \ \ \ \ \ \ \ \ \ \ \ \ \ \ \
\\
rTT^{t}+N=sI,  \ \ \ \ \ \ \ \ \ \ \ \ \ \ \ \ \ \ \ \ \ \ \ \ \ \ \ \ \ \ \  \ \ \ \ \ \ \ \ \ \ \ \ \
\\
N=sI-rTT^{t}=\left(
\begin{array}{cc}
s-r(TT^{t})_{11} & -r(TT^{t})_{12} \\
-r(TT^{t})_{21} & s-r(TT^{t})_{22}%
\end{array}\right). \
\end{eqnarray}
For any $r\geq 1$, there exists $s\geq 1$ such that Eq.(35) holds, then we get
$(TT^{t})_{11}=(TT^{t})_{22}$, $(TT^{t})_{12}=(TT^{t})_{21}=0$. Let
\begin{eqnarray}
(TT^{t})_{11}=(TT^{t})_{22}=t^{2},
\\
T=tO \text{ \ with \ } OO^{t}=I,
\end{eqnarray}
where $t$, $\alpha $, $\beta $ are all real numbers, $O$ is a real
orthogonal matrix. Taking Eqs.(36,37) into Eq.(35) we get
\begin{eqnarray}
N=\left(
\begin{array}{cc}
w & 0 \\
0 & w%
\end{array} \right),
\end{eqnarray}
where $w=s-rt^{2}$. The conditions $N\geq 0,\det N\geq (\det T-1)^{2}$ yield
\begin{eqnarray}
w\geq |t^{2}\det O-1|.
\end{eqnarray}
We conclude this section as Theorem 2 below.

\textbf{\emph{Theorem 2.}} A one-mode Gaussian operation is called incoherent if it maps any
incoherent Gaussian state into incoherent Gaussian states. Any one-mode incoherent
Gaussian operation can be expressed by $(T,N)$ in Eqs.(37-39).

\section{A coherence measure of one-mode Gaussian states based on relative entropy}
For any one-mode Gaussian state $\rho (V,d)$, we define a coherence measure
as
\begin{eqnarray}
C(\rho )=\inf_{\delta }\{S(\rho ||\delta ):\delta \text{ \  is \ an \ incoherent \ state\}},
\end{eqnarray}
where $S(\rho ||\delta )=tr(\rho \log _{2}\rho )-tr(\rho \log _{2}\delta )$
is the relative entropy, $inf$ runs over all incoherent Gaussian states. The entropy of $\rho $, $S(\rho )=-tr(\rho \log
_{2}\rho )$ is \cite{Holevo1999}
\begin{eqnarray}
S(\rho )=g(\nu )=\frac{\nu +1}{2}\log _{2}\frac{\nu +1}{2}-\frac{\nu -1}{2}%
\log _{2}\frac{\nu -1}{2},
\end{eqnarray}
where $\nu =\sqrt{\det V}$. We now calculate $\sup_{\delta }tr(\rho \log
_{2}\delta )$. Suppose
\begin{eqnarray}
\delta (\overline{n})=\sum_{n=0}^{\infty }\frac{\overline{n}^{n}}{(%
\overline{n}+1)^{n+1}}|n\rangle \langle n|,
\end{eqnarray}
then
\begin{eqnarray}
tr[\rho \log _{2}\delta ]=tr[\rho _{diag}\log _{2}\delta
]=\sum_{n=0}^{\infty }\rho _{nn}\log _{2}\frac{\overline{n}^{n}}{(\overline{n%
}+1)^{n+1}}
\nonumber \\
=(\sum_{n=0}^{\infty }n\rho _{nn})\log \overline{n}%
-(\sum_{n=0}^{\infty }n\rho _{nn}+1)\log (\overline{n}+1), \ \
\end{eqnarray}
where $\rho _{diag}=\sum_{n=0}^{\infty }\rho _{nn}|n\rangle \langle n|$. It follows that
\begin{eqnarray}
\frac{\partial }{\partial \overline{n}}tr[\rho \log _{2}\delta ]=\frac{1}{%
\ln 2}\frac{1}{\overline{n}+1}[\frac{\sum_{n=0}^{\infty }\rho _{nn}n}{%
\overline{n}}-1].
\end{eqnarray}
Let $\frac{\partial }{\partial \overline{n}}tr[\rho \log _{2}\delta ]=0,$ we
get
\begin{eqnarray}
\overline{n}=\sum_{n=0}^{\infty }\rho _{nn}n.
\end{eqnarray}
The remaining is how to calculate $\overline{n}=\sum_{n=0}^{\infty }\rho
_{nn}n.$
\begin{eqnarray}
\overline{n}=\sum_{n=0}^{\infty }\langle n|\rho |n\rangle
n=\sum_{n=0}^{\infty }\langle n|\rho a^{+}a|n\rangle =tr(\rho a^{+}a)
\nonumber \\
=\int
\frac{d^{2}\alpha }{\pi }\langle \alpha |\rho a^{+}a|\alpha \rangle =\int
\frac{d^{2}\alpha }{\pi }\alpha \langle \alpha |\rho a^{+}|\alpha \rangle  \ \
\nonumber \\
=\int \frac{d^{2}\alpha }{\pi }\alpha \int \frac{d^{2}\lambda }{\pi }\chi
(\rho ,\lambda )\langle \alpha |D(-\lambda )a^{+}|\alpha \rangle , \ \ \
\end{eqnarray}
where we have used $a^{+}a|n\rangle =n|n\rangle $, the coherent state $%
|\alpha \rangle $, $a|\alpha \rangle =\alpha |\alpha \rangle $, and Eq.(3).

It is easy to check that
\begin{eqnarray}
a^{+}|\alpha \rangle =e^{-\frac{|\alpha |^{2}}{2}}\sum_{n=1}^{\infty }\frac{%
n\alpha ^{n-1}}{\sqrt{n!}}|n\rangle ,
\\
\langle \alpha |D(-\lambda )=e^{\frac{\alpha \lambda ^{\ast }-\alpha ^{\ast
}\lambda }{2}}\langle \alpha +\lambda |,
\end{eqnarray}
thus direct algebras show that
\begin{eqnarray}
\langle \alpha |D(-\lambda )a^{+}|\alpha \rangle =(\alpha ^{\ast }+\lambda
^{\ast })e^{\frac{3\alpha \lambda ^{\ast }-\alpha ^{\ast }\lambda +|\alpha
|^{2}-|\alpha +\lambda |^{2}}{2}}. \ \
\\
\overline{n}=\frac{1}{\pi ^{2}}\int \int dx_{\alpha }dy_{\alpha
}dx_{\lambda }dy_{\lambda }[x_{\alpha }^{2}+y_{\alpha }^{2}+x_{\alpha
}x_{\lambda }+y_{\alpha }y_{\lambda }
\nonumber \\
+i(y_{\alpha }x_{\lambda }-x_{\alpha
}y_{\lambda })]\exp [-\frac{1}{2}\sum_{ij}Q_{ij}x_{i}x_{j}+%
\sum_{i}a_{i}x_{i}], \ \
\end{eqnarray}
where $\alpha =x_{\alpha }+iy_{\alpha },$ $\lambda =x_{\lambda }+iy_{\lambda
},$ $a=(0,0,-id_{2},id_{1}),$ $(x_{1},x_{2},x_{3},x_{4})=(x_{\alpha },y_{\alpha
},x_{\lambda },y_{\lambda }).$
\begin{eqnarray}
Q=\left(
\begin{array}{cccc}
0 & 0 & 0 & 2i \\
0 & 0 & -2i & 0 \\
0 & -2i & 1+V_{22} & -V_{12} \\
2i & 0 & -V_{12} & 1+V_{11}%
\end{array}\right),
\\
Q^{-1}=\left(
\begin{array}{cccc}
\frac{1+V_{11}}{4} & \frac{V_{12}}{4} & 0 & -\frac{i}{2} \\
\frac{V_{12}}{4} & \frac{1+V_{22}}{4} & \frac{i}{2} & 0 \\
0 & \frac{i}{2} & 0 & 0 \\
-\frac{i}{2} & 0 & 0 & 0%
\end{array}\right),
\\
Q^{-1}a^{t}=(\frac{d_{1}}{2},\frac{d_{2}}{2},0,0)^{t},aQ^{-1}a^{t}=0.
\end{eqnarray}
Using the result of Ref.\cite{Withers1985} with
some algebras we get
\begin{eqnarray}
\overline{n}=\frac{1}{4}(V_{11}+V_{22}+d_{1}^{2}+d_{2}^{2}-2).
\end{eqnarray}
In conclusion, we get

\begin{eqnarray}
C[\rho (V,d)] =\frac{\nu -1}{2}\log \frac{\nu -1}{2}-\frac{\nu +1}{2}\log
\frac{\nu +1}{2}
\nonumber \\
+(\overline{n}+1)\log (\overline{n}+1)-\overline{n}\log
\overline{n}, \\
\nu =\sqrt{\det V}=\sqrt{V_{11}V_{22}-V_{12}^{2}}, \ \ \ \ \ \ \ \ \ \ \ \ \ \ \ \ \ \
\\
\overline{n}=\frac{1}{4}%
(V_{11}+V_{22}+d_{1}^{2}+d_{2}^{2}-2). \ \ \ \ \ \ \ \ \ \ \ \ \ \ \ \
\end{eqnarray}

We next prove $C[\rho ]$ is nondecreasing under any incoherent operation.
For any incoherent operation $O$, suppose $C[\rho ]=S(\rho ||\overline{\rho }%
)$ with the thermal state $\overline{\rho }=\overline{\rho }(\overline{n})$
and $\overline{n}=\sum_{n=0}^{\infty }\rho _{nn}n$ as specified in Eq.(57),
then we have
\begin{eqnarray}
C[O(\rho )]\leq S[O(\rho )||O(\overline{\rho })]\leq S[\rho ||\overline{%
\rho }]=C[\rho ].
\end{eqnarray}
In Eq.(58) above, the first inequality comes from the definition of $C[O(\rho
)]$ and the fact that $O(\overline{\rho })$ is a thermal state, the second
inequality comes from the monotonicity of relative entropy under completely
positive and trace preserving mapping \cite{Lindblad1975}.

From Eqs.(55-57), we see that the coherence measure $C[\rho (V,d)]$ is strictly
monotonically decreasing in $\nu $ while strictly monotonically increasing
in $\overline{n}$. For pure Gaussian states $\nu =1$ reaches the minimum of $\nu $%
. In this sense, we say that the maximally coherent states are pure.

\section{Multi-mode Gaussian states}

We extend the results of one-mode Gaussian states into multi-mode Gaussian
states. For the positive integer $m\geq 2$, an $m$-mode Gaussian state $\rho(V,d)$ is described by \cite{Weedbrook2012} its covariance matrix $V$, a $2m\times2m$ real symmetric positive matrix, and its displacement vector $d$, a $2m$ dimensional real vector. $V$ satisfies $V+i\Omega\geq0$ with $\Omega =\left(
\begin{array}{cc}
0 & 1 \\
-1 & 0%
\end{array}\right)
^{\otimes m}.$

\subsection{Incoherent states}
For fixed orthonormal basis $(\{|n\rangle \}_{n=0}^{\infty })^{\otimes m}$
with positive integer $m\geq 2$, the diagonal states is of the form
\begin{eqnarray}
\rho ^{A_{1}A_{2}...A_{m}}  \ \ \ \ \ \ \ \ \ \ \ \ \ \ \ \ \ \ \ \ \ \ \ \ \ \ \ \  \ \ \ \ \ \ \ \ \ \ \ \ \ \ \ \ \ \ \ \ \ \ \ \ \
\nonumber \\
=\sum_{n_{1},...,n_{m}}\rho
_{n_{1}n_{1},...,n_{m}n_{m}}|n_{1}\rangle \langle n_{1}|\otimes ...\otimes
|n_{m}\rangle \langle n_{m}|,
\end{eqnarray}
where $A_{i}$ denotes the $i$th mode, and
\begin{eqnarray}
\rho _{n_{1}n_{1},...,n_{m}n_{m}} \ \ \ \ \ \ \ \ \ \ \ \ \ \ \ \ \ \ \ \ \ \ \ \ \ \ \  \ \ \ \ \ \ \ \ \ \ \ \ \ \ \ \ \ \ \
\nonumber \\
=\langle n_{1}|\langle n_{2}|...\langle
n_{m}|\rho ^{A_{1}A_{2}...A_{m}}|n_{m}\rangle ...|n_{2}\rangle |n_{1}\rangle. \ \ \ \ \ \
\end{eqnarray}
It is easy to check that
\begin{eqnarray}
[\rho ^{A_{1}A_{2}},\rho ^{A_{1}}]=0,
\end{eqnarray}
where $\rho ^{A_{1}}$ is the reduced state with respect to the $A_{1}$ mode,
$\rho ^{A_{1}A_{2}}$ is the reduced state with respect to the $A_{1}A_{2}$
modes, and $[\cdot ]$ denotes commutator. Recall that for two-mode Gaussian
states \cite{Xu2015}
\begin{eqnarray}
[\rho ^{A_{1}A_{2}},\rho ^{A_{1}}]=0\Leftrightarrow \rho ^{A_{1}A_{2}}=\rho
^{A_{1}}\otimes \rho ^{A_{2}}.
\end{eqnarray}
Together with Theorem 1 above, we get that the incoherent two-mode Gaussian
states are of the form
\begin{eqnarray}
\rho ^{A_{1}A_{2}}=\rho _{th}^{A_{1}}(\overline{n_{1}})\otimes \rho
_{th}^{A_{2}}(\overline{n_{2}}),
\end{eqnarray}
where $\rho _{th}^{A_{1}}(\overline{n_{1}})$ and $\rho _{th}^{A_{2}}(%
\overline{n_{2}})$ are all thermal states with mean numbers $\overline{n_{1}}
$ and $\overline{n_{2}}.$

For $m$-mode incoherent Gaussian state $\rho ^{A_{1}A_{2}...A_{m}}$, we
consider its covariance matrix $V$. Since $\rho ^{A_{i}A_{j}}$ for any $%
1\leq i\leq j$ is of the form $\rho ^{A_{i}A_{j}}=\rho ^{A_{i}}\otimes \rho
^{A_{j}}$, hence the covariance matrix $V$ must be of diagonal form, that is
\begin{eqnarray}
\rho ^{A_{1}A_{2}...A_{m}}=\otimes _{i=1}^{m}\rho _{th}^{A_{i}}(\overline{%
n_{i}}).
\end{eqnarray}
\subsection{Incoherent operation}

An $m$-mode Gaussian operation is described by $(T,N,\overline{d})$, it
performs on the Gaussian state $\rho (V,d)$ and get the Gaussian state with
the covariance matrix and displacement vector as \cite{Holevo2001}
\begin{eqnarray}
d\rightarrow Td+\overline{d},V\rightarrow TVT^{t}+N,
\end{eqnarray}
where $\overline{d}\in R^{2m}$, $N$, $T$ are $2m\times 2m$
real matrices satisfying
\begin{eqnarray}
N+i\Omega -iT\Omega T^{t}\geq 0.
\end{eqnarray}
Similar to the one-mode case, it is easy to determine the incoherent
operation is of the form as follows. $T$ consists of $\{t_{i}O_{i}%
\}_{i=1}^{m}$ with $t_{i}$ real number, $O_{i}$ $2\times 2$ real matrix
satisfying $O_{i}O_{i}^{t}=I,$ each $(2i-1,2i)$ row has just $t_{i}O_{i}$,
each $(2j-1,2j)$ column has just one of $\{t_{i}O_{i}\}_{i=1}^{m}$, and
other elements are all zero. $N=diag\{w_{1}I,w_{2}I,...,w_{m}I\}$ with $%
w_{i}\geq 0$ and $I$ being $2\times 2$ identity$.$The condition Eq.(65) then
reads
\begin{eqnarray}
w_{i}\geq |t_{i}^{2}\det O_{i}-1|  \text{ \ for \ all \ }  i.
\end{eqnarray}
\subsection{A coherence measure}
Similar to the one-mode case, we can generalize Eqs.(55-57) into $m$-mode case,
that is
\begin{eqnarray}
C[\rho (V,d)] =-S(\rho ) \ \ \ \ \ \ \ \ \ \ \ \ \ \ \ \ \ \ \ \ \ \ \ \ \ \ \ \ \ \ \ \ \ \ \ \ \ \ \ \ \ \ \ \ \ \
\nonumber \\
+\sum_{i=1}^{m}[(\overline{n_{i}}+1)\log _{2}(%
\overline{n_{i}}+1)-\overline{n_{i}}\log _{2}\overline{n_{i}}],\ \\
S(\rho ) =-\sum_{i=1}^{m}[\frac{\nu _{i}-1}{2}\log _{2}\frac{\nu _{i}-1}{2}%
-\frac{\nu _{i}+1}{2}\log _{2}\frac{\nu _{i}+1}{2}], \ \\
\overline{n_{i}} =\frac{1}{4}%
\{V_{11}^{(i)}+V_{22}^{(i)}+[d^{(i)}]_{1}^{2}+[d^{(i)}]_{2}^{2}-2\},\ \ \ \ \ \ \ \ \ \ \ \ \
\end{eqnarray}
where $S(\rho )$ is the entropy of $\rho $ \cite{Holevo1999}, $\{\nu _{i}\}_{i=1}^{m}$ are symplectic eigenvalues of $V
$ \cite{Weedbrook2012}, $\overline{n_{i}}$ is
determined by the $i$th-mode covariance matrix $V^{(i)}$ and displacement
vector $d^{(i)}$. We can prove Eq.(68) fulfills (C2a) in the similar way as the one-mode case.

\section{Summary}
In summary, along the line of quantifying coherence of finite-dimensional
quantum states, we provided a measure for Gaussian states. To this aim, we
proved that the incoherent Gaussian states are just thermal states. We
defined the Gaussian incoherent operations as the Gaussian operations which
maps incoherent states into incoherent states and found out the structure of
Gaussian incoherent operations. The central result is that we provided a
coherence measure for  Gaussian states based on the relative entropy, it
satisfies $(C1)$ and $(C2a)$.

There remain many questions for future investigations. Firstly, whether or
not we can establish the counterparts of $(C2b)$ and $(C3)$ for Gaussian
states, in present work we only adopt $(C1)$ and $(C2a)$ as necessary
conditions for any coherence measure. Secondly, how about the behaviors of
coherence in Gaussian dynamical systems, such as frozen coherence
\cite{Bromley2015}, sudden change etc.

This work was supported by the Chinese Universities Scientific Fund (Grant
No.2014YB029) and the National Natural Science Foundation of China (Grant
No.11347213).


\begin{thebibliography}{99}
\bibitem{Glauber1963} R. Glauber, Phys. Rev. 131, 2766 (1963).

\bibitem{Sudarshan1963} E. Sudarshan, Phys. Rev. Lett. 10, 277 (1963).

\bibitem{Luo2005} S. Luo, Theor. Math. Phys. 143, 681 (2005).

\bibitem{Aberg2006} J. \AA berg, arXiv:quant-ph/0612146.

\bibitem{Monras2014} A. Monras, A. Ch\c{e}ci\'{n}ska, and A. Ekert, New J. Phys. 16, 063041 (2014).

\bibitem{Aberg2014} J. \AA berg, Phys. Rev. Lett. 113, 150402 (2014).

\bibitem{Girolami2014} D. Girolami, Phys. Rev. Lett. 113, 170401 (2014).

\bibitem{Baumgratz2014} T. Baumgratz, M. Cramer, and M. B. Plenio, Phys. Rev. Lett. 113, 140401 (2014).


\bibitem{Bennett1996} C. H. Bennett, H. J. Bernstein, S. Popescu, and B. Schumacher, Phys. Rev. A 53, 2046 (1996).

\bibitem {Vedral1998} V. Vedral and M. B. Plenio, Phys. Rev. A 57, 1619 (1998).

\bibitem{Plenio2007} M. B. Plenio and S. Virmani, Quantum Info. Comput. 7, 1 (2007).

\bibitem{Horodecki2009} R. Horodecki, P. Horodecki, M. Horodecki, and K. Horodecki, Rev. Mod. Phys. 81, 865 (2009).

\bibitem{Streltsov2015} A. Streltsov, U. Singh, H. S. Dhar, M. N. Bera, and G. Adesso, Phys. Rev. Lett. 115, 020403 (2015).

\bibitem{Shao2015} L.-H. Shao, Z. Xi, H. Fan, and Y. Li, Phys. Rev. A 91, 042120 (2015).

\bibitem{Bromley2015} T. R. Bromley, M. Cianciaruso, and G. Adesso, Phys. Rev. Lett. 114, 210401 (2015).

\bibitem{Xi2015} Z. Xi, Y. Li, and H. Fan, Sci. Rep. 5, 10922 (2015).

\bibitem{Pires2015} D. P. Pires, L. C. C\'{e}leri, and D. O. Soares-Pinto, Phys. Rev. A 91, 042330 (2015).

\bibitem{Bera2015} M. N. Bera, T. Qureshi, M. A. Siddiqui, and A. K. Pati, Phys. Rev. A 92, 012118 (2015).

\bibitem{Singh2015} U. Singh, M. N. Bera, H. S. Dhar, and A. K. Pati, Phys. Rev. A 91, 052115 (2015).

\bibitem{Yadin2015} B. Yadin and V. Vedral, arXiv:1505.03792.

\bibitem{Yuan2015} X. Yuan, H. Zhou, Z. Cao, and X. Ma, Phys. Rev. A 92, 022124 (2015).

\bibitem{Yao2015} Y. Yao, X. Xiao, L. Ge, and C. P. Sun, Phys. Rev. A 92, 022112 (2015).

\bibitem{Du2015} S. Du, Z. Bai, and Y. Guo, Phys. Rev. A 91, 052120 (2015).



\bibitem{Braunstein2005} S. L. Braunstein and P. van Loock, Rev. Mod. Phys. 77, 513 (2005).

\bibitem{Wang2007} X. B. Wang, T. Hiroshimab, A. Tomitab, and M. Hayashi, Phys. Rep. 448, 1 (2007).

\bibitem{Weedbrook2012} C. Weedbrook, S. Pirandola, R. Garc\'{\i}a-Patr\'{o}n, N. J. Cerf, T. C. Ralph, J. H. Shapiro, and S. Lloyd, Rev. Mod. Phys. 84, 621 (2012).



\bibitem{Isserlis1918} L. Isserlis, Biometrika 12, 134 (1918).

\bibitem{Wick1950} G. C. Wick, Physical Review 80 (2), 2689 (1950).

\bibitem{Withers1985} C. S. Withers Bull. Austral. Math. Soc. 32, 103 (1985).



\bibitem{Holevo2001} A. S. Holevo, and R. F. Werner, 2001, Phys. Rev. A 63, 032312(2001).

\bibitem{Holevo1999} A. S. Holevo, M. Sohma, and O. Hirota, Phys. Rev. A 59, 1820(1999).

\bibitem{Lindblad1975} G. Lindblad, Commun. Math. Phys. 40, 147 (1975).


\bibitem{Xu2015} J. Xu. Int. J. Theor. Phys. 54, 860 (2015).
\end{thebibliography}
\end{document}